\documentclass[12pt]{iopart}
\usepackage{iopams}
\usepackage{setstack}

\newcommand{\cP}{\ensuremath{\mathcal{P}}}
\newcommand{\cT}{\ensuremath{\mathcal{T}}}

\begin{document}

\title[Does the complex deformation of the Riemann equation exhibit shocks?]
{Does the complex deformation of the Riemann equation exhibit shocks?}

\author[Bender and Feinberg]{Carl~M~Bender\footnote{Permanent address:
Department of Physics, Washington University, St. Louis MO 63130, USA;
\\{\footnotesize{\tt email: cmb@wustl.edu}}} and
Joshua~Feinberg\footnote{Permanent address: Department of Physics,
University of Haifa at Oranim, Tivon 36006, Israel and
Department of Physics, Technion, Haifa 32000, Israel;
\\{\footnotesize{\tt email: joshua@physics.technion.ac.il}}}}

\address{Center for Nonlinear Studies, Los Alamos National Laboratory,
Los Alamos, NM 87545, USA}

\date{today}

\begin{abstract}
The Riemann equation $u_t+uu_x=0$, which describes a one-dimensional
accelerationless perfect fluid, possesses solutions that typically develop
shocks in a finite time. This equation is $\cP\cT$ symmetric. A one-parameter
$\cP\cT$-invariant complex deformation of this equation, $u_t-iu(iu_x)^\epsilon=
0$ ($\epsilon$ real), is solved exactly using the method of characteristic
strips, and it is shown that for real initial conditions, shocks cannot develop
unless $\epsilon$ is an odd integer.
\end{abstract}

\submitto{\JPA}

\section{Introduction}
\label{s1}

The concept of $\cP\cT$-symmetric quantum mechanics was introduced in 1998
\cite{r1}. Since then, there have been many studies of the properties of such
quantum theories \cite{r2,r3,r4,r5,r6}. In an effort to understand the
underlying mathematical structure of $\cP\cT$-symmetric quantum mechanics, there
have recently been a number of studies of classical differential equations that
are symmetric under $\cP\cT$ reflection. These studies include an examination of
the complex solutions of the differential equations for the classical-mechanical
systems associated with $\cP\cT$-symmetric quantum-mechanical systems
\cite{r2,r7,r8,r9}.
Additional studies have been made of various $\cP\cT$-symmetric
classical-mechanical systems such as the $\cP\cT$-symmetric pendulum \cite{r10},
the $\cP\cT$-symmetric Lotka-Volterra equation \cite{r11,r12}, and the $\cP
\cT$-symmetric Euler equations that describe free rigid-body rotation
\cite{r12}.

Many well-known nonlinear wave equations exhibit $\cP\cT$ symmetry. For example,
the Korteweg-de Vries and generalized Korteweg-de Vries equations \cite{r13},
the Camassa-Holm equation \cite{r14}, the Sine-Gordon and Boussinesq equations
\cite{r13} are all $\cP\cT$ symmetric. Once it is known that a differential
equation is $\cP\cT$ symmetric, it is possible to introduce a continuous
parameter $\epsilon$ to deform the equation into the complex domain in such a
way as to preserve the $\cP\cT$ symmetry. Such a deformation gives rise to large
parametric classes of new equations whose properties exhibit an interesting
dependence on $\epsilon$. For example, in recent papers a complex family of
nonlinear wave equations obtained by deforming the Korteweg-de Vries equation
was examined \cite{r15,r16}.

In this paper we study a simplified version of the complex class of Korteweg-de
Vries equations considered in Ref.~\cite{r15}; namely, the first-order nonlinear
wave equations obtained by removing the dispersive third-order-derivative term.
This class of equations is the $\cP\cT$-symmetric deformation of the Riemann
equation. The virtue of studying this simpler class of equations is that they
can be solved {\it exactly and in closed form} for all values of the deformation
parameter $\epsilon$, as we show in this paper. This allows us to determine how
the properties of the complex solution depend on $\epsilon$.

The well known and heavily studied Riemann equation
\begin{equation}
\label{e1}
u_t+uu_x=0
\end{equation}
describes a one-dimensional incompressible accelerationless fluid. The velocity
profile $u(x,t)$ represents the velocity of the fluid at the point $x$ and at
time $t$. For the initial condition
\begin{equation}
\label{e2}
u(x,0)=f(x)
\end{equation}
the exact solution to (\ref{e1}) can be written in implicit form as
\begin{equation}
\label{e3}
u=f(x-ut).
\end{equation}

The Riemann equation (\ref{e1}) is symmetric under $\cP\cT$ reflection, where
$\cP$ represents spatial reflection and $\cT$ represents time reversal. (Note
that under $\cP\cT$ reflection $u$ does not change sign because it is a
velocity.) A simple way to deform the Riemann equation into the complex domain
in such a way as to preserve its $\cP\cT$ invariance is to introduce the real
parameter $\epsilon$ as follows:
\begin{equation}
\label{e4}
u_t-iu(iu_x)^\epsilon=0.
\end{equation}
Note that as in quantum mechanics, we assume that $\cT$ is an antilinear
operator that has the effect of reversing the sign of the complex number $i$.
For the special value $\epsilon=1$ this equation reduces to the conventional
Riemann equation (\ref{e1}). We will examine the solutions to (\ref{e4}) that
arise from the initial condition in (\ref{e2}).

Let us recall how to solve the standard Riemann equation in (\ref{e1}). The
Riemann equation is a typical example of a {\it quasilinear} partial
differential equation (one that is linear in the partial derivatives) that can
be solved using the method of characteristics. To do so we parametrize the
initial condition ({\ref{e2}) in terms of the parameter $r$:
\begin{equation}
\label{e5}
t=0,\quad x=r,\quad u=f(r).
\end{equation}
Next, we construct the system of three coupled ordinary differential equations
known as the {\it characteristic differential equations}:
\begin{equation}
\label{e6}
{dt\over ds}=1,\quad{dx\over ds}=u,\quad{du\over ds}=0.
\end{equation}
In general, when the characteristics for a quasilinear equation are solved
subject to the initial conditions imposed at $s=0$, the resulting solution has
the general form $x=x(s,r)$, $t=t(s,r)$, $u=u(s,r)$. Carrying out this procedure
for the case of the Riemann equation produces the solution
\begin{equation}
\label{e7}
t=s,\quad x=f(r)s+r,\quad u=f(r).
\end{equation}
The last step is to eliminate the variables $r$ and $s$ in favor of the original
variables $x$ and $t$, which yields the solution $u(x,t)$ of (\ref{e1}) given
implicitly in (\ref{e3}).

It is well known that for most choices of the initial condition $f(x)$ the
solution $u(x,t)$ eventually becomes a multiple-valued function of $x$. This
happens because the characteristic curves in the $x-t$ plane for the Riemann
equation are the straight lines $x=f(x_0)t+x_0$, as we can see from (\ref{e7}).
These lines are in general not all parallel because $f$ is not a constant, and
thus some of the characteristics may eventually cross. It is physically
unacceptable for the velocity field of a fluid to be multiple-valued. To avoid
multiple-valued solutions, it is conventional to require that when two
characteristic lines cross, a shock develops. A {\it shock} is a jump
discontinuity as a function of $x$ in the fluid profile $u(x,t)$. A solution
$u(x,t)$ containing a shock is not differentiable at the position of the shock
and it is called a {\it weak solution} \cite{r13}.

The purpose of this paper is to show that deforming the Riemann equation into
the complex domain has the effect of softening the shock singularities, as was
conjectured in Ref.~\cite{JF}. Specifically, we will show in the next section
that shocks can only form when $\epsilon$ takes odd-integer values. To show this
we will solve the deformed Riemann equation (\ref{e4}) exactly using the method
of characteristic strips, which is explained in the Appendix.

\section{Shock Formation Time for the $\cP\cT$-Deformed Riemann Equation}

To find out whether the $\cP\cT$-deformed Riemann equation (\ref{e4}) develops
shocks, we examine the problem quantitatively: We calculate the precise amount
of time required for a shock to form. To demonstrate the technique, we
begin by calculating the shock-formation time for the standard Riemann equation
in (\ref{e1}).

A shock forms as soon as the coordinate transformation in (\ref{e7}) becomes
singular. To determine when this happens we calculate the Jacobian of this
transformation
\begin{eqnarray}
\label{e8}
J={\partial(t,x)\over\partial(s,r)}=\det\,\left[\begin{array}{cc} 1~ & 0 \\
f(r)~ & sf'(r)+1\end{array}\right]=1+sf'(r).
\end{eqnarray}
This transformation ceases to be invertible as soon as $J$ vanishes, namely, at
the {\em minimal positive} value of $s=-1/f'(r)$. Thus, the shock-formation time
$t_{\rm shock}$ is given by
\begin{equation}
\label{e9}
t_{\rm shock}={{\rm positive~min}\atop{x\in I\!\! R}}\,[-1/f'(x)].
\end{equation}

For $\epsilon\neq 1$, (\ref{e4}) is nonlinear, and not quasilinear, and the
method of characteristics does not apply. Instead, we shall solve (\ref{e4})
using the method of {\em characteristic strips}, a generalization of the method
of characteristics, which in some cases can be used to solve nonlinear
first-order partial differential equations.

The method of characteristic strips is explained briefly in the Appendix. Using
the notation in the Appendix, we rewrite (\ref{e4}) as $F(x,y,u,p,q)=0$, where
$t\equiv y$, $p=u_x$, $q=u_t$, and
\begin{equation}
\label{e10}
F=q-iu(ip)^\epsilon.
\end{equation}
The procedure is to seek a coordinate transformation of the form $x=x(s,r)$ and
$t=t(s,r)$ such that the five arguments of $F$ satisfy the five
characteristic-strip equations in (\ref{A18}), namely,
\begin{equation}
\label{e11}
{dx\over ds}=-\epsilon q/p,\quad{dt\over ds}=1,\quad{du\over ds}=(1-\epsilon)q,
\quad{dp\over ds}=pq/u,\quad{dq\over ds}=q^2/u.
\end{equation}
Here we have substituted (\ref{e10}) into (\ref{A18}) and also substituted $F=0$
in various places, after taking the appropriate derivatives. Note that for
$\epsilon=1$, the equations for $x$, $t$, and $u$ agree with the characteristic
differential equations (\ref{e6}) for the Riemann equation (\ref{e1}).

Remarkably, the coupled system of ordinary differential equations (\ref{e11})
admits a closed explicit solution:\footnote{It is not always possible to use the
method of characteristic strips to solve a nonlinear partial differential
equation because the technical assumption that is made in (\ref{A16}) may not
hold.}
\begin{eqnarray}
\label{e12}
p &=& G(D-\epsilon s)^{-1/\epsilon},\quad q=E(D-\epsilon s)^{-1/\epsilon},
\nonumber\\
u&=&E(D-\epsilon s)^{(\epsilon-1)/\epsilon},\quad t=s+H,\quad x=-\epsilon
Es/G+I,
\end{eqnarray}
where $D$, $E$, $G$, $H$, and $I$ are functions of the parameter $r$ and are yet
to be determined from the initial conditions (\ref{e2}).

Next, as explained in the Appendix, we verify that (\ref{e12}), which solves the
characteristic-strip equations (\ref{e11}), constitutes a solution of the
partial differential equation (\ref{e4}). To do so we impose the condition that
$F(x,t,u,p,q)=0$ on the five functions in (\ref{e12}). In fact, this condition
is nontrivial, and it translates into the condition
\begin{equation}
\label{e13}
(iG)^\epsilon=-i,
\end{equation}
which determines $G$. [For irrational values of $\epsilon$ the solution to this
equation has infinitely many branches, and we choose the branch that tends to
$-1$ as $\epsilon\to1$. This is the solution that corresponds to the solution of
the Riemann equation (\ref{e1}) as $\epsilon\to1$.]

Next, we rewrite the initial conditions $u(x,0)=f(x)$ in parametric form. On the
initial curve at $s=0$ we have
\begin{equation}
\label{e14}
x=r,\quad t=0,\quad u=f(r).
\end{equation}
Since $u=u(x,t)$, we have $u_r=p{dx\over dr}+q{dt\over dr}$. Also ${du\over dr}=
f'(r)$, ${dx\over dr}=1$, ${dt\over dr}=0$. Thus,
\begin{equation}
\label{e15}
u_r=p(r)=f'(r).
\end{equation}
Also, since $F=0$, we deduce from (\ref{e10}) that
\begin{equation}
\label{e16}
q(r)=iu(r)[ip(r)]^\epsilon=if(r)[if'(r)]^\epsilon.
\end{equation}
We impose the initial conditions by setting $s=0$ in (\ref{e12}). Then, using
(\ref{e15}) and (\ref{e16}), we obtain
\begin{eqnarray}
\label{e17}
p(0)&=&f'(r)=GD^{-1/\epsilon},\quad q(0)=if(r)[if'(r)]^\epsilon=ED^{-1/\epsilon}
,\nonumber\\
u(0)&=&f(r)=ED^{{\epsilon-1\over\epsilon}},\quad t(0)=0=H,\quad x(0)=r=I.
\end{eqnarray}

From the first two equations in (\ref{e17}) we solve for $D$ and $E$ in terms of
$f(r)$, $f'(r)$, and $G$. Substituting these expressions for $D$ and $E$ in the
third equation in (\ref{e17}) gives an equation for $G$ that coincides with
(\ref{e13}). Hence, we obtain a self-consistent result. Substituting $G$ into
the expressions for $D$ and $E$ gives
\begin{equation}
\label{e18}
D=-i[if'(r)]^{-\epsilon},\quad E/G=-f(r)[if'(r)]^{\epsilon-1}.
\end{equation}

From the last two equations in (\ref{e17}) we have
\begin{eqnarray}
\label{e19}
I=r,\quad H=0,
\end{eqnarray}
and gathering our results from (\ref{e12}), (\ref{e18}), and (\ref{e19}), we
finally obtain
\begin{equation}
\label{e20}
t=s,\quad x=r+\epsilon f(r)[if'(r)]^{\epsilon-1}s,\quad u=f(r).
\end{equation}
Surprisingly, even though (\ref{e4}) is nonlinear, we have solved it in {\it
closed form}. [Note that for $\epsilon=1$ this solution reduces to the solution
of the Riemann equation (\ref{e1}).]

The next step in the application of the method of characteristic strips is to
solve for $r$ and $s$ in terms of $x$ and $t$, and to substitute the expression
for $r$ into $u(r)$ to obtain the desired solution $u(x,t)$. This can be done if
and only if the Jacobian
\begin{equation}
\label{e21}
J={\partial(t,x)\over\partial(s,r)}=\det\,\left[
\begin{array}{cc}{\partial t\over\partial s}~& {\partial t\over\partial r}\\
{\partial x\over\partial s}~&{\partial x\over\partial r}\end{array}\right]=
{\partial x\over\partial r}=1+\epsilon s{d\over dr}\left\{f(r)[if'(r)]
^{\epsilon-1}\right\}
\end{equation}
does not vanish.

Clearly, for {\em real} $f(r)$, that is, for real initial conditions, $J$ never
vanishes unless $\epsilon$ is an {\it odd integer}. This establishes the
principal result of this paper. When $\epsilon$ is an odd integer, the Jacobian
may vanish, and a shock will form at time $t_{\rm shock}$ given by
\begin{equation}
\label{e22}
t_{\rm shock}={{\rm positive~min}\atop{r\in I\!\! R}}\,
\left(-1\over{\epsilon{d\over dr}\left\{f(r)[if'(r)]^{\epsilon-1}\right\}}
\right).
\end{equation}

\subsection{Case $\epsilon=2$:}

For $\epsilon=2$ (\ref{e4}) can be reduced to the Riemann equation for a {\it
pure-imaginary} velocity field as follows: Consider (\ref{e4}) at $\epsilon=2$
and multiply it by $u$. Define $v=u^2/2$. The resulting equation is $v_t=-i
v_x^2$. Take the $x$ derivative of both sides and define $w=iv_x=iuu_x$.
Clearly, for real $u(x,0)=f(x)$, $w(x,0)=if(x)f'(x)$ is pure-imaginary and
$w$ satisfies the Riemann equation $w_t+2ww_x=0$. The Jacobian (\ref{e21}) in
this case is $J=1+2is{d\over dx}[f(x)f'(x)]$; $J$ can never vanish for real
$f(x)$, in accordance with our general result.

\subsection{Case $\epsilon=3$:}
According to our general result, shocks may form at $\epsilon=3$.
In this case (\ref{e22}) reads
\begin{equation}
\label{e23}
t_{\rm shock}={{\rm positive~min}\atop{r\in I\!\! R}}\,\left({1\over3{d\over dr}
\left\{f(r)[f'(r)]^2\right\}}\right).
\end{equation}

As a concrete example, we take the initial velocity profile to be the Cauchy
distribution $f(x)=1/(x^2+1)$. For this case, (\ref{e23}) becomes
\begin{equation}
\label{e24}
t_{\rm shock}={{\rm positive~min}\atop{x\in I\!\! R}}\,\left[\frac{(1+x^2)^6}
{24x(1-4x^2)}\right].
\end{equation}
This function has two maxima and two minima, which are located at
$x=\pm0.216\,621$ and at $x=\pm0.769\,392$. The lower positive minimum
occurs at $x=0.216\,621$, and at this value of $x$, it equals $0.311\,791$.
Thus, this initial condition evolves into a shock at $t=0.311\,791$.

\section*{Appendix: Concise Summary of the Method of Characteristic Strips}

In this Appendix we give a concise recipe for using the method of characteristic
strips to solve nonlinear first-order partial differential equations. This
Appendix is brief, and we remark that a complete discussion of the method of
characteristic strips would be rather long and complicated (see, for example,
Refs.~\cite{r17,r18}).

We wish to solve the general first-order partial differential equation
\begin{equation}
\label{A1}
F(x,y,u,u_x,u_y)=0
\end{equation}
for the unknown function $u(x,y)$ subject to the initial condition $u(x,0)=f(x
)$. [Thus, $u_x(x,0)=f'(x)$, while $u_y(x,0)$ is determined by the equation $F=
0$ at $y=0$.] To this end we introduce two more unknowns, $p$ and $q$ and
rewrite (\ref{A1}) as
\begin{equation}
\label{A2}
F(x,y,u,p,q)=0,
\end{equation}
subject to the constraints
\begin{equation}
\label{A3}
u=u(x,y),\quad p=u_x=p(x,y),\quad q=u_y=q(x,y).
\end{equation}
In order to solve this equation in a manner similar to the method of
characteristics, we seek a coordinate transformation of the form $x=x(s,r)$,
$y=y(s,r)$, which will have the effect of reducing (\ref{A1}) to a closed system
of coupled {\it ordinary} differential equations in the variable $s$. The
initial condition will be parametrized in terms of $r$.

We can immediately write down eight differential equations associated with the
partial differential equation (\ref{A1}):
\begin{equation}
\label{A4}
p=u_x,
\end{equation}
\begin{equation}
\label{A5}
q=u_y,
\end{equation}
\begin{equation}
\label{A6}
p_y=q_x,
\end{equation}
\begin{equation}
\label{A7}
{du\over ds}=p{dx\over ds}+q{dy\over ds},
\end{equation}
\begin{equation}
\label{A8}
{dp\over ds}=p_x{dx\over ds}+ p_y{dy\over ds},
\end{equation}
\begin{equation}
\label{A9}
{dq\over ds}=q_x{dx\over ds}+q_y{dy\over ds},
\end{equation}
\begin{equation}
\label{A10}
F_x+F_up+F_pp_x+F_qq_x=0,
\end{equation}
\begin{equation}
\label{A11}
F_y+F_uq+F_pp_y+F_qq_y=0.
\end{equation}
It is easy to check that (\ref{A10}) and (\ref{A11}) lead to ${dF\over ds}=0$,
which is consistent with (\ref{A2}).

Next, we use (\ref{A6}) to rewrite (\ref{A10}) as
\begin{equation}
\label{A12}
F_x+F_up+F_pp_x+F_qp_y=0,
\end{equation}
and similarly, we rewrite (\ref{A11}) as
\begin{equation}
\label{A13}
F_y+F_uq+F_pq_x+F_qq_y=0.
\end{equation}
Since our goal is to find expressions for the parametric derivatives ${dx\over
ds}$, ${dy\over ds}$, ${du\over ds}$, ${dp\over ds}$, and ${dq\over ds}$ in
terms of $x$, $y$, $u$, $p$, and $q$, we must try to eliminate all reference to
the higher derivatives $p_x$, $p_y$, $q_x$, and $q_y$. We proceed as follows:
Using (\ref{A8}) and (\ref{A9}), we eliminate $p_y$ and $q_x$ as
\begin{equation}
\label{A14}
p_y=\left({dp\over ds}-p_x{dx\over ds}\right)\bigg/{dy\over ds}\quad{\rm and}
\quad q_x=\left({dq\over ds}-q_y{dy\over ds}\right)\bigg/{dx\over ds}.
\end{equation}
Substituting these expressions into (\ref{A11}) and (\ref{A12}) we obtain
\begin{eqnarray}
\label{A15}
F_x+F_u p+p_x\left(F_p-F_q{dx\over ds}\bigg/{dy\over ds}\right)+F_q{dp\over
ds}\bigg/{dy\over ds} &=& 0,\nonumber\\
F_y+F_uq+q_y\left(F_q-F_p{dy\over ds}\bigg/{dx\over ds}\right)+F_p{dq\over
ds}\bigg/{dx\over ds} &=& 0.
\end{eqnarray}

It is now clear that the only way to obtain a closed system of five coupled
ordinary differential equations is to impose the additional requirement that
\begin{equation}
\label{A16}
F_p{dy\over ds}=F_q{dx\over ds}.
\end{equation}
Upon imposing (\ref{A16}), we see that (\ref{A11}) and (\ref{A12}) assume the
particularly simple form
\begin{equation}
\label{A17}
F_x+F_up+F_q{dp\over ds}\bigg/{dy\over ds}=0,\quad
F_y+F_uq+F_p{dq\over ds}\bigg/{dx\over ds}=0.
\end{equation}
We emphasize that the requirement (\ref{A16}) is not necessarily valid, and
after solving the partial differential equation (\ref{A1}) it is necessary to
verify the correctness of (\ref{A16}).

Next, we make use of the functional degree of freedom to {\it reparametrize};
that is, to introduce a new parameter $s'=g(s)$, which is an arbitrary function
$g(s)$ of the old parameter $s$. This freedom allows us to choose ${dx\over ds}=
F_p$, and from (\ref{A16}) it follows that ${dy\over ds}=F_q$. In this way we
obtain the desired $s$ derivatives as
\begin{eqnarray}
\label{A18}
{dx\over ds}&=&F_p,\quad{dy\over ds}=F_q,\quad{du\over ds}=pF_p+qF_q,\nonumber\\
{dp\over ds}&=&-F_x-F_up,\quad{dq\over ds}=-F_y-F_uq.
\end{eqnarray}
These are the characteristic-strip equations. These are a generalization of the
three coupled characteristic ordinary differential equations for quasilinear
partial differential equations to the five characteristic-strip equations for
fully nonlinear partial differential equations. However, we emphasize that there
is a new aspect; namely, that after solving the system (\ref{A18}), it is
necessary to verify that the assumption made in (\ref{A16}) is valid.

The problem of solving the nonlinear equation (\ref{A1}) is now converted to
solving the coupled system (\ref{A18}). Once this is done and (\ref{A16})
has been verified, it is then necessary to impose the initial conditions, and
this will introduce the dependence on the parameter $r$. The final step is to
eliminate $r$ and $s$ in favor of the original variables $x$ and $y$, and
thereby to obtain the solution $u=u(x,y)$.

\vspace{0.5cm}
\footnotesize
\noindent
As an Ulam Scholar, CMB receives financial support from the Center for Nonlinear
Studies at the Los Alamos National Laboratory. CMB is also supported by a grant
from the U.S. Department of Energy. JF thanks the Center for Nonlinear Studies
for its hospitality and for partial financial support.

\end{document}